\documentclass[12pt]{iopart}
\usepackage{amssymb}
\usepackage{color}
\usepackage{graphicx}
\usepackage{appendix}
\usepackage{float}
\usepackage{hyperref}
\usepackage{iopams}
\expandafter\let\csname equation*\endcsname\relax

\expandafter\let\csname endequation*\endcsname\relax

\usepackage{amsmath}
\input{Qcircuit}
\usepackage[tight,FIGBOTCAP]{subfigure}
\usepackage[matrix,frame,arrow]{xypic}

\begin{document}

\title{Overcoming erasure errors with multilevel systems}

\author{Sreraman Muralidharan$^{1,3}$, Chang-Ling Zou$^{2,3}$, Linshu Li$^{2,3}$, Jianming Wen$^{2,3}$, Liang Jiang$^{2,3}$}

\address{$^1$Department of Electrical Engineering, Yale University, New Haven, CT 06511 USA}
\address{$^2$Department of Applied Physics, Yale University, New Haven, CT 06511 USA}
\address{$^3$Yale Quantum Institute, Yale University, New Haven, CT 06520, USA}

\ead{liang.jiang@yale.edu}
\vspace{10pt}
\begin{indented}
\item[]December 2016
\end{indented}

\begin{abstract}
We investigate the usage of highly efficient error correcting codes of multilevel systems to protect encoded quantum information from erasure errors and implementation to repetitively correct these errors. Our scheme makes use of quantum polynomial codes to encode quantum information and generalizes teleportation based error correction for multilevel systems to correct photon losses and operation errors in a fault-tolerant manner. We discuss the application of quantum polynomial codes to one-way quantum repeaters. For various types of operation errors, we identify different parameter regions where quantum polynomial codes can achieve a superior performance compared to qubit based quantum parity codes. 

\end{abstract}

\section{Introduction}

A quantum erasure channel replaces a qubit (qudit) with an ``erasure state'' that is orthogonal to all the basis states of a qubit (qudit) with a certain probability, thereby erasing the qubit (qudit) and enabling the receiver know that it has been erased \cite{Bennett1997}. Physically, erasure errors may occur in various situations, such as leakage to other states \cite{Preskill1998, Wu2002, Fazio1999}, atom losses \cite{Vala2005}, and photon losses \cite{Knill2001, Duan2004, Kok2007, Bell2014}.
For ion-trap systems, leakage processes occur when the qubit moves out of the idealized two-level sub-space to a larger space \cite{Preskill1998, Wu2002, Fazio1999}. For quantum memories with optical lattices, back ground gas collisions can eject the atoms leading to atom losses  \cite{Vala2005}
Photon losses occur in linear optical quantum computing schemes \cite{Knill2001} due to absorption in optical interconnects or optical fiber. Undoubtedly, protecting quantum information from erasure errors is a significant challenge for practical quantum computation and long distance quantum communication.

Specifically, for long distance quantum communiction through optical fibers, photon losses lead to an exponential penality in resources and time. The exponential penality can be overcome by establishing intermediate repeater stations and actively  correcting for erasure and operation errors at these stations. Three generations of quantum repeaters have been proposed based on the different approaches used to correct erasure and operation errors \cite{Muralidharan2015a}.  The first generation employs heralded entanglement generation between neighboring repeater stations to correct erasure errors and entanglement  purification to correct operation errors \cite{Briegel98}. Heralded entanglement generation needs two-way classical communication between neighboring repeater stations, while entanglement purification needs two-way classical communication between remote repeater stations. The second generation \cite{Jiang2008, Bratzik14, Epping2016, Pant2016}  employs heralded entanglement generation to correct erasure errors and quantum error correction to correct operation errors. Quantum error correction doesn't require any form of two-way classical communication. The third generation uses quantum error correction to correct both loss and operation errors, and avoids any form of two-way classical communication between repeater stations, thereby rendering ultrafast communication over transcontinental distances \cite{Fowler2010,Munro2012a, Muralidharan2014,Ewert2015,Namiki2016, glaudell2016}. Since erasure errors are actively corrected in these repeater schemes, it is crucial to investigate quantum error correcting codes that can correct erasure errors very efficiently \cite{Varnava2006, Barrett2010, Stace2010, Gingrich2003}. So, far only quantum parity codes (QPC) have been optimized for third generation quantum repeaters \cite{Muralidharan14, Munro2012a}.  

Due to the quantum no-cloning theorem \cite{Bennett1997}, no error correcting code can correct erasure errors deterministically when the erasure rate is above $50\%$. 
There have been significant advances in searching for quantum codes that can correct up to $50\%$ erasure error rates.
Varnava \textit{et. al.} \cite{Varnava2006} showed that by using tree-like cluster states for encoding, one can correct erasure errors when erasure rate is close to the $50\%$ bound for one-way quantum computation. Stace and Barrett \cite{Stace2010, Barrett2010} demonstrated that surface codes can also correct erasure errors when error rates are close to $50\%$. However, qubit based codes often require large code size to enable the correction of a large fraction of erasure errors. 

Quantum error correcting codes of higher-dimensional systems provide a promising alternative to qubit based encoding schemes to correct erasure errors. For example, quantum polynomial codes (QPyC) \cite{Aharonov2008, Cleve1999} are a class of CSS codes that were introduced in the context of fault-tolerant quantum computation \cite{Aharonov2008} and shown to be useful for constructing threshold quantum secret sharing schemes \cite{Cleve1999}. One can encode a secret qudit into $2k+1$ qudits (with prime dimension $d\geq 2k+1$) and distribute one qudit to each of the many parties, so that at least $(k+1)$ of them should get together to reconstruct the secret. This makes the $[[2k+1,1,k+1]]_d$ QPyC a good choice for the correction of erasure errors up to a fraction of $k/(2k+1) \rightarrow 50\%$ for a large $k$.

This paper is organized as follows: We first perform a comparison between $[[3,1,2]]_3$ code and
the $[[4,2,2]]$ code. In section III, we investigate the ability of general QPyC to correct erasure
errors and compare them to surface codes. In section IV, we show that QPyC can be used for third generation (or one-way) quantum repeaters. We also compare our quantum repeater scheme with other schemes based on quantum parity codes \cite{Muralidharan14, Munro2012a} in the presence of operation errors. Here, we identify the parameter regimes where QPyC performs better than QPC. The price for this improved performance against erasure errors is that more complex multi-mode operations must be implemented for encoding and readout operations. In section V, we discuss the key experimental techniques needed for the physical implementation of our scheme and provide potential experimental procedures based on atom mediated photonic gates for multilevel systems.

\section{Three-qutrit code vs Four-qubit code}
To illustrate how the error correction for erasure errors works, consider the four-photon $[[4,2,2]]$ code that maps two qubits into the logical states \cite{Gingrich2003, Lu2008}, 
\begin{eqnarray}
|00\rangle_L = &\frac{1}{2} (|00\rangle+|11\rangle)_{12}(|00\rangle+|11\rangle)_{34},\nonumber \\
|01\rangle_L = &\frac{1}{2} (|00\rangle-|11\rangle)_{12}(|00\rangle-|11\rangle)_{34},\\
|10\rangle_L = &\frac{1}{2} (|01\rangle+|10\rangle)_{12}(|01\rangle+|10\rangle)_{34},\nonumber \\
|11\rangle_L = &\frac{1}{2} (|01\rangle-|10\rangle)_{12}(|01\rangle-|10\rangle)_{34},\nonumber
\end{eqnarray}
where the subscripts 1-4 denote qubits 1-4. Here, $|0\rangle$ and $|1\rangle$ corresponds to a single photon occupying two different modes such as polarization or time-bin states '01' and '10' respectively. The [[4,2,2]] code requires a total of four pairs of modes (eight modes) and four photons. The loss of a photon leads to the vacuum state of the associated pair of modes, which corresponds to an erasure error. Suppose we transmit the encoded state through a channel and it undergoes one photon loss 
(i. e. erasure of one of qubits), then the quantum error correcting code enables the reconstruction of the encoded state as follows. First, we use quantum non-demolition measurement to extract the total excitation number for each pair of modes (without destroying the qubit), which will be one if there is no photon loss, or zero if the photon is lost. Hence, the photon loss can be identified as an erasure error of the associated qubit. For example, if the first qubit is erased, to reconstruct the encoded state we apply two CNOT gates betweeen qubits 3 and 2, and qubits 4 and 2 respectively. Then, we measure qubit 2 in the Z basis and 
and use the measurement outcome to reconstruct the logical state. This can be seen by studying the logical operators of the $[[4,2,2]]$ code, $X_L^{(1)} = IXIX$, $X_L^{(2)} = IZIZ$, $Z_L^{(1)} =IIZZ$ and  $Z_L^{(2)} = IIXX$. After the CNOT gates, the logical operators are transformed into $IIIX$, $IZZI$, $IIZZ$ and $IIXX$ respectively. After the gates, a Z-measurement on the second qubit is needed to decode the $[[4,2,2]]$ code. We refer the readers to Ref. \cite{Gingrich2003}, where an 
alternative non-destructive method has been proposed to recover the logical qubits.
It has been proven that we need at least four qubits to correct one erasure error \cite{Grassl1997}. 

Alternatively, one can also correct an erasure error using a $[[3,1,2]]_3$ code which encodes a logical qutrit into three physical qutrits as \cite{Cleve1999}, 
\begin{eqnarray}
&|0\rangle_L& = \frac{1}{\sqrt{3}} ( |000\rangle + |111\rangle + |222\rangle)_{123},\\
&|1\rangle_L& = \frac{1}{\sqrt{3}} ( |012\rangle + |120\rangle + |201\rangle)_{123}, \nonumber \\
&|2\rangle_L& = \frac{1}{\sqrt{3}} ( |021\rangle + |102\rangle+ |210\rangle)_{123}, \nonumber 
\label{eq:threequtrit}
\end{eqnarray}
where the subscripts 1-3 represent qutrits 1-3. Here, we note that each qutrit represents different time bin states and not photon number. More specifically, $|0\rangle$, $|1\rangle$ and $|2\rangle$ represent temporal modes '001', '010' and '100' respectively. So, the $[[3,1,2]]_3$ code requires a
total of three triplets of modes (nine modes) and three photons. Photon loss leads to the vacuum state of the corresponding triplet of modes, which can identified as the erasure error of the associated qutrit.
If qutrit is erased, it is possible to reconstruct the encoded qutrit by performing an addition modulo 3 operation between the other two qutrits. For example, for the incoming state
$\alpha|0\rangle_L+\beta|1\rangle_L+\gamma|2\rangle_L$, if the first qutrit is erased, then the state of second and third qutrits is given by
\begin{align}
&\rho_{23} = \sum_{i=0}^2 |\psi_i\rangle_{23}\langle\psi_i|,
\intertext{where,} 
&|\psi_0\rangle = \alpha|00\rangle_{23}+\beta|12\rangle_{23}+\gamma|21\rangle_{23} \nonumber \\
&|\psi_1\rangle = \alpha|11\rangle_{23}+\beta|20\rangle_{23}+\gamma|02\rangle_{23}\nonumber \\
&|\psi_2\rangle = \alpha|22\rangle_{23}+\beta|01\rangle_{23}+\gamma|10\rangle_{23}\nonumber. 
\end{align}
Using two addition modulo 3 operations\footnote{An addition modulo 3 operation refers to a SUM gate here is described in detail in the section IV.} between qutrits 2 and 3, and then between qutrits 3 and 2, we can reconstruct the encoded state. 

\begin{figure}[h]
\begin{center}
\subfigure[]{
\label{fig:compare}
\includegraphics[width=7cm]{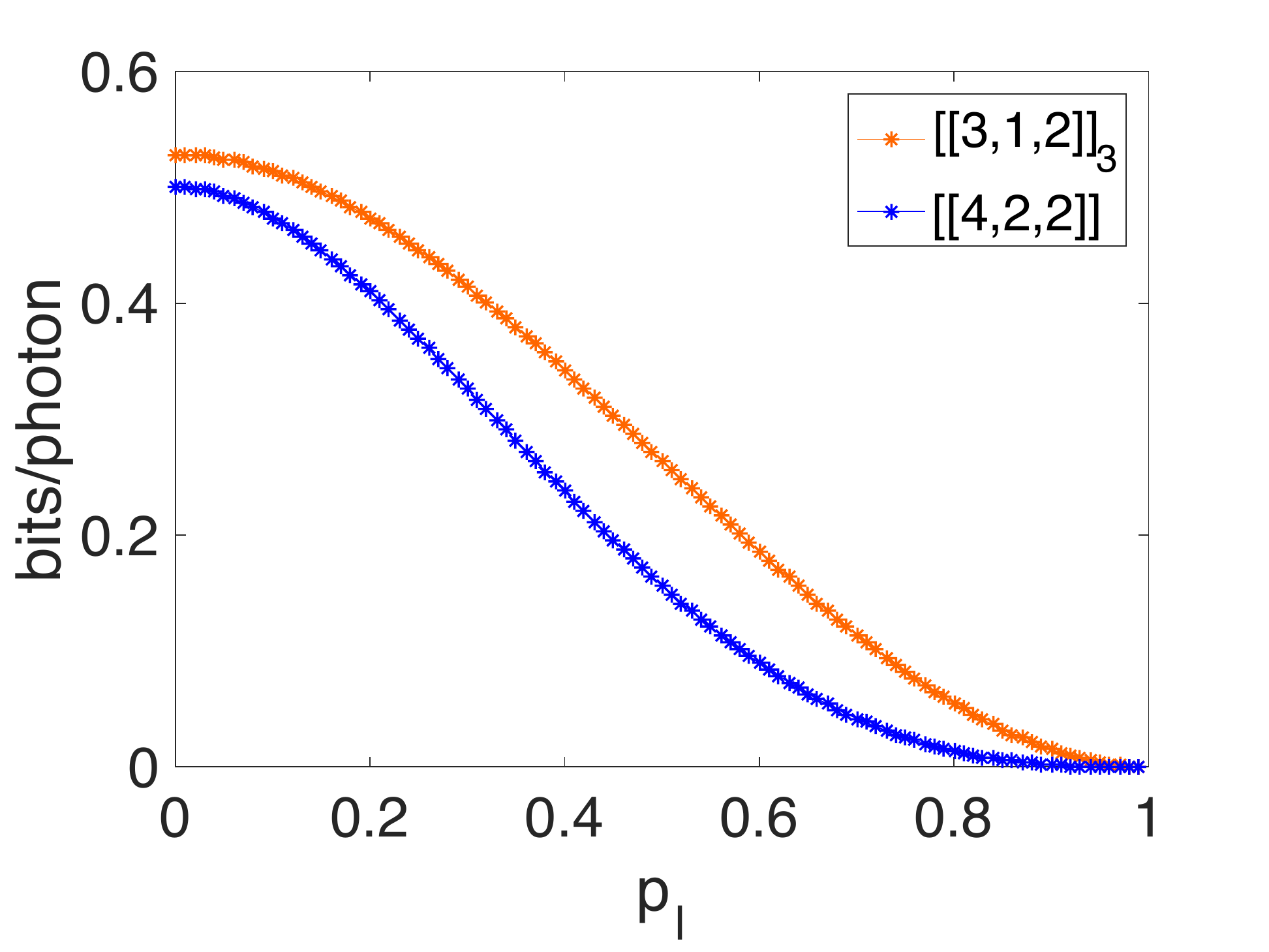}
}
\subfigure[]{
\label{fig:effectiveerror}
\includegraphics[width=7cm]{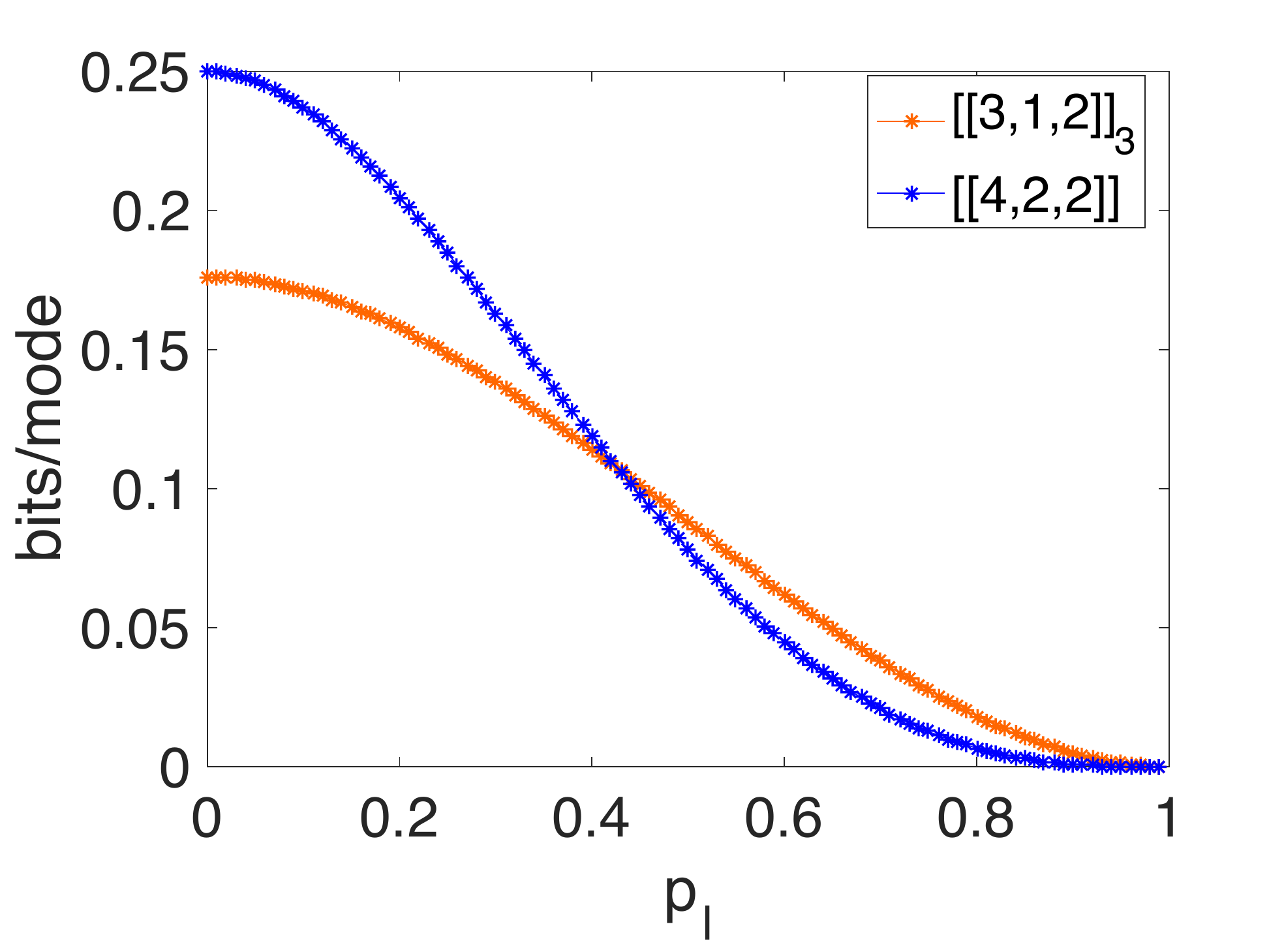}
}
\caption[fig:fa1]{a) A comparison between bits/photon that can be achieved with 
$[[3,1,2]]_3$ code and $[[4,2,2]]$ code respectively. b) A comparison between bits/mode that can be achieved with $[[3,1,2]]_3$ code and $[[4,2,2]]$ code respectively.
}
\end{center}
\end{figure}

We now compare the $[[3,1,2]]_3$ code and the $[[4,2,2]]$ code for different loss rate per photon, $p_l$.  Since each logical qutrit can carry $\mbox{log}_{2}3$ qubits of information, we multiply a factor of $\mbox{log}_{2}3$ when computing bits/photon and bits/mode for $[[3,1,2]]_3$ code. As illustrated in Fig. \ref{fig:compare}, the bits/photon is always higher for the $[[3,1,2]]_3$ code compared to the $[[4,2,2]]$ code for all loss rates $p_l$. With regard to the bits/mode, for large loss rates $(p_l>42\%)$, the $[[3,1,2]]_3$ code performs better than the $[[4,2,2]]$ code. In principle, by concatenating the $[[3,1,2]]_3$ code, it is possible to supress more erasure errors. However, to be resource efficient we will consider the generalization of the $[[3,1,2]]_3$ codes - QPyC in the forthcoming section.

\section{Quantum Polynomial codes} 
The qutrit $[[3,1,2]]_3$ code can be generalized to $d$-level ($d$ is a prime number) qudit system as a $[[2k+1,1,k+1]]_d$ code, which encodes one logical qudit into $2k+1$ physical qudits and can correct up to $k$ erasure errors \cite{Cleve1999}. The QPyC code is a CSS code which can be obtained from 
classical $[2k+1,1,k+1]$ polynomial code. The encoded states of the QPyC code are given by \cite{Aharonov2008, Cleve1999},
\begin{equation}
|s\rangle_L = \sum_{{c_{k}=s},|c\in \textbf{F}^{k+1}} |p(\textbf{c},x_0), p(\textbf{c},x_1),..., p(\textbf{c},x_{2k})\rangle.
\end{equation}
Here, $s$ labels the logical states with  $s \in \{0,1,2...d-1\}$. The labels $x_0$,$x_1$,...,$x_{2k}$ are $2k+1$ distinct elements of $\textbf{F}$, where $\textbf{F} =Z_d$. 
$d$ is the first prime number that is no less than $2k+1$. The polynomial is given by $p(\textbf{c},t) = \sum_{j=0}^{k} c_j t^j $ with $\textbf{c}=(c_0,c_1,...c_{k}) \in \textbf{F}^{k+1}$. For $k=1$, $d=3$, it can be seen that 
\begin{equation}
|s\rangle_L = \sum_{\substack {c_{0}=0\\\\c_1=s }}^2 |c_0\rangle |c_0 + c_1\rangle |c_0 + 2 c_1\rangle, 
\end{equation}
yielding the $[[3,1,2]]_3$ code (see Eq. (\ref{eq:threequtrit})). The success probability of recovering the encoded quantum information after it has undergone erasure errors is given by
\begin{equation}
\text{P}^{\text{QPyC}}_{\text{success}} = \sum_{j=0}^{k} {2k+1 \choose j} {p_l}^{j} {(1-p_l)}^{2k+1-j}.
\label{eq:success}
\end{equation}
Note that for a reasonable code size, it is possible to achieve substantially low (e. g. $10^{-5}$) failure rate $\text{P}_\text{fail}=1-\text{P}^{\text{QPyC}}_{\text{success}}$ even in the presence of high erasure rate $p_l$. Note that with about 40 qudits one can suppress the failure rate to $10^{-6}$ for $20\%$ erasure rate. 

It can be seen from Eq. (\ref{eq:success}) that at $p_l = 50\%$, $\text{P}^{\text{QPyC}}_{\text{success}} = 1/2$ independent of the code size. Further, note that the success probability of error correction
has a phase transition behaviour with $k \rightarrow \infty$, $\text{P}^{\text{QPyC}}_{\text{success}} \rightarrow 1$ for $p_l< 50\%$ and $\text{P}^{\text{QPyC}}_{\text{success}} \rightarrow 0$ for $p_l>50\%$. 
One can obtain the critical exponent of phase transition by noting that for $k \rightarrow \infty$, around the fixed point $\text{P}^{\text{QPyC}}_{\text{success}} (k,\frac{1}{2})=\frac{1}{2}$, the asymptotic
expression is
\begin{small}
\begin{eqnarray}
&\text{P}^{\text{QPyC}}_{\text{success}}(k,p_l) = 1-\left(\begin{array}{c}
2k+1\\
k
\end{array}\right)2^{-2k-1}\frac{\sqrt{\pi}\Gamma(k+2)}{2\Gamma(k+3/2)}\nonumber \\
& - 2^{-2k-1}\frac{4k{}_{2}F_{1}(2,1-k;k+3;-1)}{k+2}(\frac{1}{2}-p_l)+o[(p_l-\frac{1}{2})^{2}].
\end{eqnarray}
\end{small}
For $k\rightarrow\infty$, making the approximation that $\left(\begin{array}{c}
2k+1\\
k
\end{array}\right)\sim\sqrt{\frac{1}{k\pi}}2^{2k+1}$, $\frac{\Gamma(k+2)}{\Gamma(k+3/2)}\approx\sqrt{k}$ and $\frac{_{2}F_{1}(2,1-k;k+3;-1)}{k+2}\sim\frac{1}{2}$,
then 
\begin{eqnarray}
\text{P}^{\text{QPyC}}_{\text{success}}(k,p_l) & \approx & \frac{1}{2}-2\sqrt{\frac{k}{\pi}}(\frac{1}{2}-p_l).
\end{eqnarray}
That means for given $\text{P}^{\text{QPyC}}_{\text{success}}$,
\begin{equation}
k\approx\frac{\pi}{4}\left(\frac{\text{P}^{\text{QPyC}}_{\text{success}}-\frac{1}{2}}{\frac{1}{2} - p_l}\right)^{2},
\end{equation}
yielding a critical exponent of 2. 

We will now compare the performance of QPyC with surface codes \cite{Stace2010} for the correction of erasure errors. Surface codes consists of a $D\times D$ square lattice with $2D^2$ qubits in total, where each qubit located on the edges of the square lattice. The logical operators $X_L$ ($Z_L$) of the surface codes are given by the product of $X$ ($Z$) operators along a non-trivial homological cycle connecting
the boundaries \cite{Stace2010}. Therefore, the success probability of bond percolation in the square lattice and its dual lattice \cite{Varnava2006} is essentially the same as the probability of decoding the
surface code. It is well known that the bond percolation threshold is $50\%$ for a square lattice, meaning that as the distance of the lattice $D\rightarrow \infty$, $\text{P}^{\text{SC}}_{\text{success}} \rightarrow 1$ for $p_l<50\%$ and $\text{P}^{\text{SC}}_{\text{success}} \rightarrow 0$ for $p_l>50\%$. 

\begin{figure}[h]
\begin{center}
\subfigure[]{
\label{fig:qpy}
\includegraphics[width=7cm]{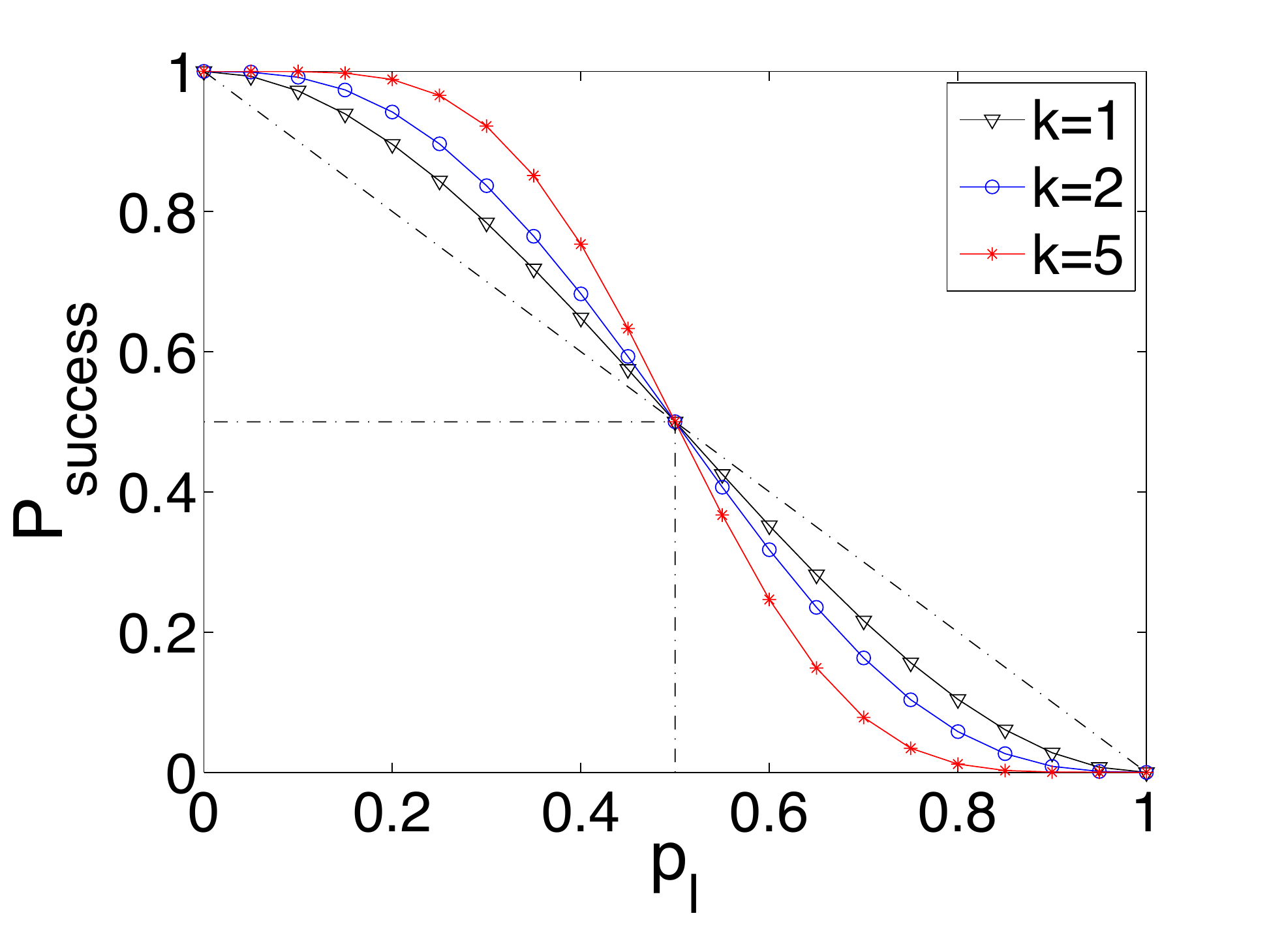}
}
\subfigure[]{
\label{fig:sc}
\includegraphics[width=6.75cm]{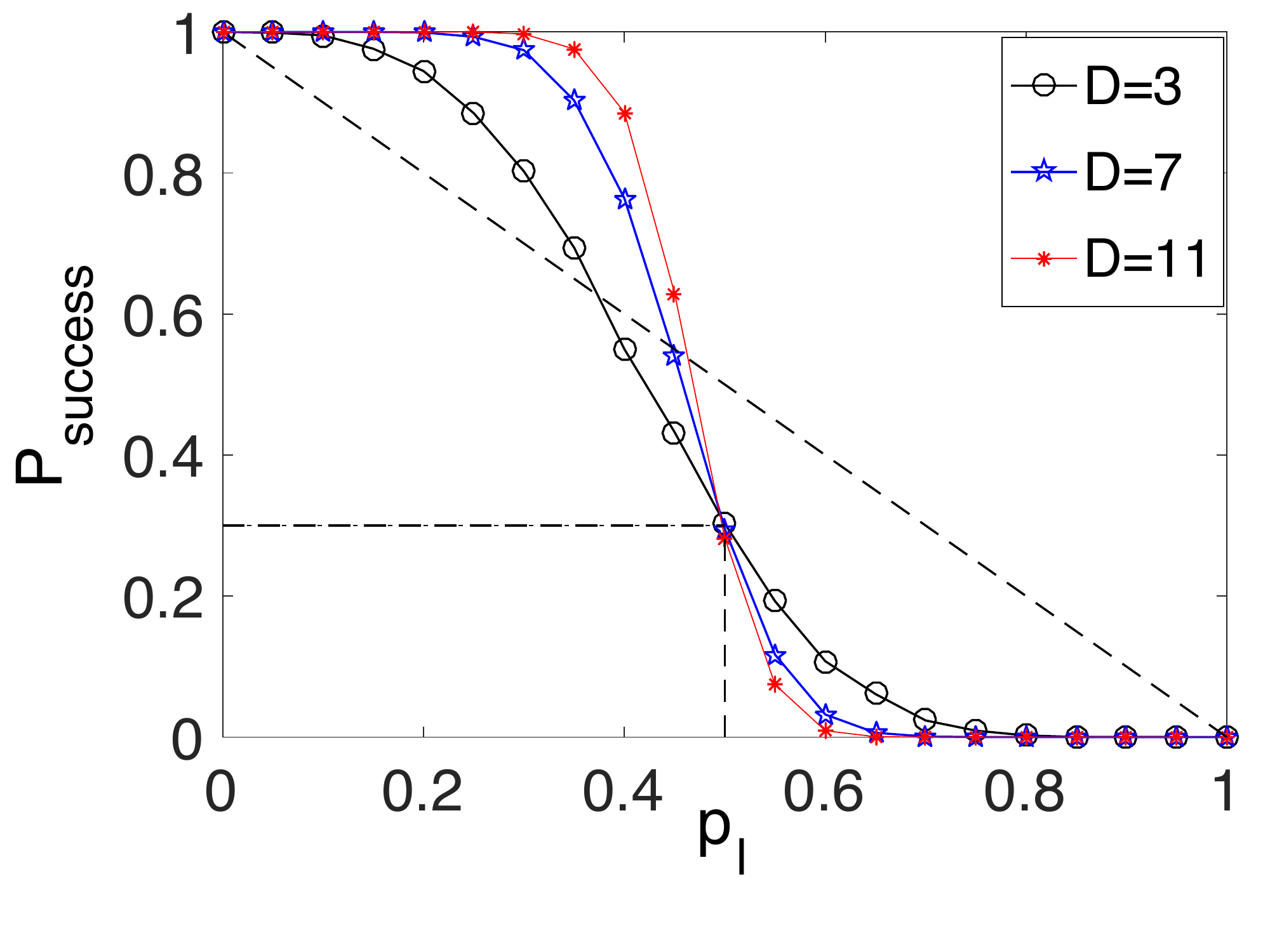}
}
\caption[fig:fa1]{(a) Success probability of QPyC for different erasure rates. (b) Success probability of decoding surface codes with three different distances (3,7,11).}
\end{center}
\end{figure}

In Fig. \ref{fig:sc} we study the success probabilities of surface codes with three different distances $D=(3,7,11)$ and find that
at $p_l=50\%$, $\text{P}^{\text{SC}}_{\text{success}} \approx 0.30$, while in Fig. \ref{fig:qpy} we see that $\text{P}^{\text{QPyC}}_{\text{success}}= 0.5$ for all code sizes. 
Therefore, when $p_l$ is around $50\%$, QPyC of any code size always outperforms surface codes. Further, it can be seen in Fig. \ref{fig:sc} that 
threshold for surface codes with $D=3$ is about $37\%$, and the threshold approaches $50\%$ for larger code sizes, while the threshold of QPyC is $50\%$ for all code sizes. In practice, 
we might operate the codes at a loss probability much lower than $50\%$. Hence, we assume 
$p_l \approx 20\%$ and compare the surface code and QPyC with $2^{2D^2} \approx (2k+1)^{(2k+1)}$ so that they have a similar Hilbert space dimension for physical encoding. As illustrated in Table 1, for smaller code sizes surface codes performs better than QPyC and 
with $D\geq 9$ and $k\geq15$, QPyC outperforms surface codes with a smaller failure probability.

\begin{table}[tbp] \centering%
\begin{tabular}{c|c|c|c}
\hline\hline
$D$ & $k $ & {Surface code} &{QPyC} \\ \hline
$5$ & $6$ & $0.0068$ & $0.007$ \\
$7$ & $9$ & $9.37\times10^{-4}$ & $0.0016$ \\
$9$ & $15$ & $1.2\times10^{-4}$ & $8.8\times10^{-5}$ \\
$11$ & $21$ & $1.3 \times10^{-5}$ & $5.23\times10^{-6}$ \\
 \hline\hline
\end{tabular}%
\caption{A comparison between the failure probabilities of error correction ($1-P_{\text{success}}$) for surface codes and QPyC with a similar size of the Hilbert space for $20\%$ erasure rates. All values were calculated for surface codes with a total of $10^6$ runs.}\label{tab:resource}%
\end{table}%

\section{Application in quantum repeaters}

One-way (or third generation) quantum repeaters \cite{Fowler2010, Munro2012a, Muralidharan14, Brinew2, Muralidharan2015a} rely on quantum error correction to relay data from one repeater station to the next. At each station, error correction operations are performed before the message is transmitted to the next station. Recently, it has been shown that QPC \cite{Ralph2005, Munro2012a, Muralidharan14} (a generalization of the $[[4,1,2]]$ code) can correct erasure errors efficiently by teleportation based error correction (TEC). Since QPyC corrects a larger fraction of erasure errors, it is reasonable to consider QPyC instead of QPC for one-way quantum repeaters, and expect a significant improvement for resource consumption and key generation rates. 

\subsection{Implementation of error correction for erasure errors}
\begin{figure}
\begin{center}
\mbox{ \ \ \ \ \ \ \ \ }
\mbox{ 
\Qcircuit @C=.1em @R=.2em @! {
\lstick{\ket{\psi}_L} & \qw & \ctrl{1} &\measuretab{X_L} & \control \cw\\
\lstick{\ket{0}_L} & \gate{S} & \gate{S} & \measuretab{Z_L} & \cwx\\
\lstick{\ket{+}_L} & \ctrl{-1} & \qw & \gate{X} \cwx & \gate{Z} \cwx & \rstick{\ket{\psi}_L} \qw\\\\
}}

\caption{TEC circuit for multilevel systems, where erasure and operation (X and Z) errors in $|\psi\rangle_L$ are corrected. The states $|0\rangle_L$ and $|+\rangle_L = \sum_{j=0}^{d-1} |j\rangle$ should be prepared fault-tolerantly and free from erasure errors. "S" refers to an encoded SUM gate
that acts on logical qudits $|i\rangle_L$ and $|j\rangle_L$ as $|i\rangle_L |j\rangle_L \rightarrow |i\rangle_L |(i+j) \mbox{mod} \ d\rangle_L$. The encoded SUM gate has pairwise implementation for CSS codes.}
\label{Fig:obt1}
\end{center}
\end{figure}
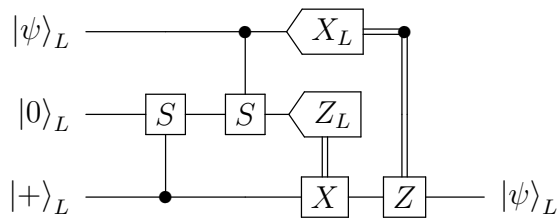

For qubit encoding schemes, it has been shown that TEC \cite{Knill2005a, Knill2005} is an effective approach for the correction of erasure and operation errors.
Since teleportation requires $X$ and $Z$ measurements on qubits, error correction for erasure and operation errors is possible if we can reliably measure the logical operators $X_L$ and $Z_L$ of the error correcting code \cite{Muralidharan14}. 
For example, the $[[4,2,2]]$ code has stabilizers $XXXX$ and $ZZZZ$ and the  following logical operators: $X_L^{(1)} = IXIX,XIXI$, $X_L^{(2)} = IZIZ,ZIZI$, $Z_L^{(1)} = ZZII,IIZZ$ and 
$Z_L^{(2)} = IIXX,XXII$. If any qubit undergoes an erasure error, all the $X_L$ operators and 
$Z_L$ operators can be measured in the TEC circuit and the encoded qubits can be retrieved. For example, if the first qubit is lost, one can calculate the operators $X_L^{(1)} = IXIX$, $X_L^{(2)} = IZIZ$, $Z_L^{(1)} = IIZZ$ and $Z_L^{(2)} = IIXX$ and retrive the encoded qubits. We generalize TEC to multilevel systems in Fig. \ref{Fig:obt1} by using generalized Pauli matrices that act on $d$-level system as $X^l|j\rangle = |j+l\rangle$ and $Z^l|j\rangle = \omega^{(lj)} |j\rangle$, $0\leq i,j\leq d-1$ \cite{Gottesman1999} and SUM gate that acts on a control qudit $|i\rangle$ with a target qudit $|j \rangle$ to produce the transformation $|i\rangle|j\rangle \rightarrow |i\rangle |(i+j) \ \mbox{mod} \ d\rangle$ \cite{Gottesman1999}. Consider the $[[3,1,2]]_3$ code with stabilizers $XXX$, $ZZZ$ and $X_L$ operator that corresponds to one of the operators $IXX^2, XX^2I, X^2IX$ and $Z_L$ operator that corresponds to one of the operators $IZ^2Z, ZIZ^2, Z^2ZI$. It can be readily seen that in the presence of an erasure error on any qutrit one can still measure the corresponding logical operator and reconstruct the encoded state. 

\subsection{Error model for operation errors} 

Suppose that each physical qudit is encoded into a $[[2k+1,1,k+1]]_d$ QPyC and is transmitted through repeater stations, where the TEC is implemented. We assume independent errors acting on physical qudits and consider an extensive error model with the following types of errors acting on the encoded qudit as follows:
\begin{enumerate}
\item The photon arrives at each repeater station with probability $1-p_l = e^{-\frac{L_0}{L_{\mbox{att}}}}$, where $L_0$ is the repeater spacing and $L_{\mbox{att}} = 20\mbox{km}$ is the attenuation length of the fiber.
\item The photon undergoes depolarization with a probability $\epsilon_d$. For simplicity, we assume that
this is also the probability of error on each physical qudit of the encoded Bell pair prepared at the repeater station. 
\item The photon experiences an additional dephasing in the matter qubit-photon coupling with a probability $\epsilon_p$.
\item Each SUM gate acting between the encoded Bell pair and the incoming qubit (see Fig. \ref{Fig:obt1}) fails at a probability $\epsilon_g$.
\end{enumerate}
By further assuming the same probabilities for all types of depolarization errors and gate errors, the transmission channel with the incoming single qudit state $A$ takes the form, 
\begin{eqnarray}
E_{c}(\rho_{A}) = (1-p_l) (1 - \epsilon_d - \epsilon_p) \rho_A + p_l |vac\rangle \langle vac| 
+\frac{(1-p_l) \epsilon_d}{d^2} \sum_{i,j=0}^{d-1} (X^i Z^j) \rho_A {(X^i Z^j)}^\dagger \nonumber \\
+\frac{(1-p_l) \epsilon_p}{d} \sum_{k=0}^{d-1} ( Z^k) \rho_A {(Z^k)}^\dagger.
\end{eqnarray}
Similarly, the imperfect gate between the incoming qudit A and qudit B at repeater station can be modeled as
\begin{eqnarray}
E_{g}(\rho_{AB})= (1-\epsilon_g)U_{sum}\rho_{AB}U^{\dagger}_{sum} \nonumber \\
+ \frac{\epsilon_g}{d^4} \sum_{a,b,c,e=0}^{d-1}(X^a Z^{b}X^{c}Z^{e}) \rho_{AB} (X^{a\dagger} Z^{b\dagger} X^{c\dagger} Z^{e\dagger}).
\end{eqnarray}

For each qudit the locally prepared logical states $|0\rangle_L$ and $|+\rangle_L = \sum_{j=0}^{d-1}|j\rangle$, we assume it undergoes depolarization, which can be modelled as
\begin{eqnarray}
E_{p}(\rho_{B}) = (1 - \epsilon_d)\rho_{B} + \frac{\epsilon_d}{d^2} \sum_{i,j=0}^{d-1} (X^i Z^j) \rho_B {(X^i Z^j)}^\dagger.
\end{eqnarray}
In contrast to the error models in \cite{Muralidharan14, Muralidharan2015a}, we track the errors in the preparation of the logical Bell state needed for TEC. From these error channels, we can calculate the probability of having an error in any one of the measurements up to the first order as
\begin{eqnarray}
&&\epsilon_X = \frac{3\epsilon_g}{d^4} (d^4-d^3) + \frac{4\epsilon_d}{d^2} (d^2-d) \nonumber \\
&&\epsilon_Z= \frac{3\epsilon_g}{d^4} (d^4-d^3) + \frac{4\epsilon_d}{d^2} (d^2-d) + \frac{\epsilon_p}{d}(d-1).
\end{eqnarray}
Once the encoded state $[[2k+1,1,k+1]]_d$ is transmitted from one repeater station to its neighbor, there exists three possibilities at the receiving repeater station
\begin{enumerate}
\item More than $k$ photons are lost in transit and the outcome of the measurement cannot be found leading to a heralded failure with probability $P_{\mathrm{fail}} =1-\text{P}^{\text{QPyC}}_{\text{success}}$. 
\item At least $k$ photons are received, but the encoded state is not decoded correctly due to the  presence of many operation errors and the encoded state is not decoded correctly with probability $P_{incorrect}$.
\item At least $k$ photons are received to make an encoded $X/Z$ measurement and the encoded state is decoded correctly with probability $P_{correct}$.
\end{enumerate}
The probability that a heralded failure does not happen at any one of the $r$ repeater stations is given by,
${[\text{P}^{\text{QPyC}}_{\text{success}}]}^r$. Let us suppose that $n_1$ photons are lost before the destination. Among the rest of the $(2k+1-n_1)$ photons that reach the destination, $n_2$ photons suffer operation errors. As such, the code can correct up to $n_1 + 2n_2 \leq k$ errors. The probability of successfully measuring the encoded X/Z measurement outcomes is given by,
\begin{small}
\begin{eqnarray}
P_{\mathrm{correct(X/Z)}} = \sum_{n_1=0}^k \sum_{n_2=0} ^{\lfloor \frac{k}{2} - \frac{n_1}{2} \rfloor} {2k+1 \choose n_1} { 2k+1-n_1 \choose n_2} \times \nonumber \\ {(p_l)}^{n_1} 
\epsilon_{\mathrm{X/Z}} ^{n_2} {(1-p_l)}^{2k+1-n_1} {(1-\epsilon_{\mathrm{X/Z}} )} ^{2k+1-n_1-n_2} 
\label{eq:pcorrect}
\ \ 
\end{eqnarray}
\end{small}
The probability of incorrect decoding of the qudit is given by,
\begin{small}
\begin{flalign}
&P_{\mathrm{incorrect(X/Z)}} = \sum_{n_1=0}^k \sum_{n_2=\lceil \frac{k}{2} - \frac{n_1}{2} + \frac{1}{2} \rceil } ^{2k+1-n_1} {2k+1 \choose n_1} {2k+1-n_1 \choose n_2} \times \nonumber \\ &{(p_l)}^{n_1} \epsilon_{\mathrm{X/Z}} ^{n_2} {(1-p_l)}^{2k+1-n_1} {(1-\epsilon_{\mathrm{X/Z}} )} ^{2k+1-n_1-n_2} \ \ 
\end{flalign}
\label{Eq:pincorrect}
\end{small}
It is easy to verify that $P_{fail(X/Z)} + P_{\mathrm{correct(X/Z)}} + P_{\mathrm{incorrect(X/Z)}} =1$.
By making a pessimistic assumption that an effective logical error in any one of the repeater stations is 
an overall logical error at the receiver's end, the logical error rate of the encoded quantum information can be defined conditioned on the success of receving enough photons as,
\begin{equation}
Q_{\mathrm{X/Z}} = 1-\frac{{\left[P_{\mathrm{correct(X/Z)}}\right]}^{r}} {{[\text{P}^{\text{QPyC}}_{\text{success}}]}^r},
\end{equation}
For the two basis protocol for quantum key distribution (where information is encoded in only two logical bases $X^1$ and $Z^1$), the asymptotic normalized secure key generation rate is \cite{Sheridan2010}. 
\begin{equation}
R = \frac{{[\text{P}^{\text{QPyC}}_{\text{success}}]}^r} {t_0} \left( \mbox{log}_{2} \mathrm{d} - 2h(Q)\right),
\end{equation}
where $t_0$ is the time taken for local operations with
\begin{eqnarray}
&Q& = \left(\frac{Q_X+Q_Z}{2}\right) \nonumber \\ 
&h(Q)& = -Q\mathrm{log}_2 \frac{Q}{d-1} - (1-Q) \mathrm{log}_2 (1-Q). 
\end{eqnarray}
\par
In Fig. 4(a), we study the dependence of the secure key generation rate $Rt_0$, with respect to the total distance of communication for small encoded blocks of QPyC, assuming 1 km repeater spacing and no operation errors. For $t_0 = 1\,\mathrm{\mu s}$, 3 qudits are sufficient to reach 700 km with key generation rates $R\approx$ 10kHz. We can increase the total range of communication to
10,000km by using 7 qudits and obtain a key generation rate $R\approx$ 1000kHz.
As the operation errors involved with the error correction $\epsilon_g=\epsilon_d=\epsilon_p=\epsilon$ increases, $P_{\mathrm{incorrect(X/Z)}}$ increases exponentially after a certain distance $L_{tot}$ and we can expect a quick decay of $R.t_0$ over a certain $L_{tot}$. The maximum $Q$ that the QPyC can tolerate depends on the dimension $d$ of the code. We can check that $Q_{\mathrm{max}} \approx 0.15, 0.21, 0.237$ for the $[[3,1,2]]_3$ code, $[[5,1,3]]_5$ code and $[[7,1,4]]_7$ codes respectively for the codes to yield non zero key generation rates. Taking only the leading terms into account in Eq. (\ref{eq:pcorrect}), we can approximate $P_{\mathrm{incorrect(X/Z)}} \approx 3\epsilon_{\mathrm{X/Z}} {(1-p_l)}^3$ for the $[[3,1,2]]_3$ code, $P_{\mathrm{incorrect(X/Z)}} \approx 4\epsilon_{\mathrm{X/Z}} p_l {5 \choose 1} {(1-p_l)}^4$ for the $[[5,1,3]]_5$ code and $P_{\mathrm{incorrect(X/Z)}} \approx 5\epsilon_{\mathrm{X/Z}} (p_l)^2 {7 \choose 2} {(1-p_l)}^5$ for the $[[7,1,4]]_7$ code respectively.  Further, since $P_{\mathrm{fail}} $ for 1km spacing is negligible, we should have $P_{incorrect} \times \frac{L_{tot}}{L_0} = Q_{\mathrm{max}}$, yielding
\begin{equation}
L_{tot,max} \approx \frac{2Q_{\mathrm{max}}L_0}{P_{incorrect(X)}+P_{incorrect(Z)}} 
\end{equation}
For example, for $\epsilon = 10^{-4}$, $L_0$ = 1km, we get $L_{\mathrm{tot, max}} = 120\text{km}, 440\text{km},1900\text{km}$ for $[[3,1,2]]_3$, $[[5,1,3]]_5$ and $[[7,1,4]]_7$ codes respectively
as confirmed by the "$\times$" lines in Fig. (4). In Fig. 4(b-d), we study the variation of $R.t_0$ for varying operation errors $\epsilon_g=\epsilon_d=\epsilon_p= 10^{-6},10^{-5},10^{-4}$ for $[[7,1,4]]_7$, $[[5,1,3]]_5$ and $[[3,1,2]]_3$ codes respectively.  We fit the points obtained from a rigorous theoretical calculation with the approximate $P_{\mathrm{incorrect}}$ mentioned above (shown as gray curve in Fig. 4) and observe a good match between them. 
\begin{figure}[h]
\begin{center}
\includegraphics[width=14cm]{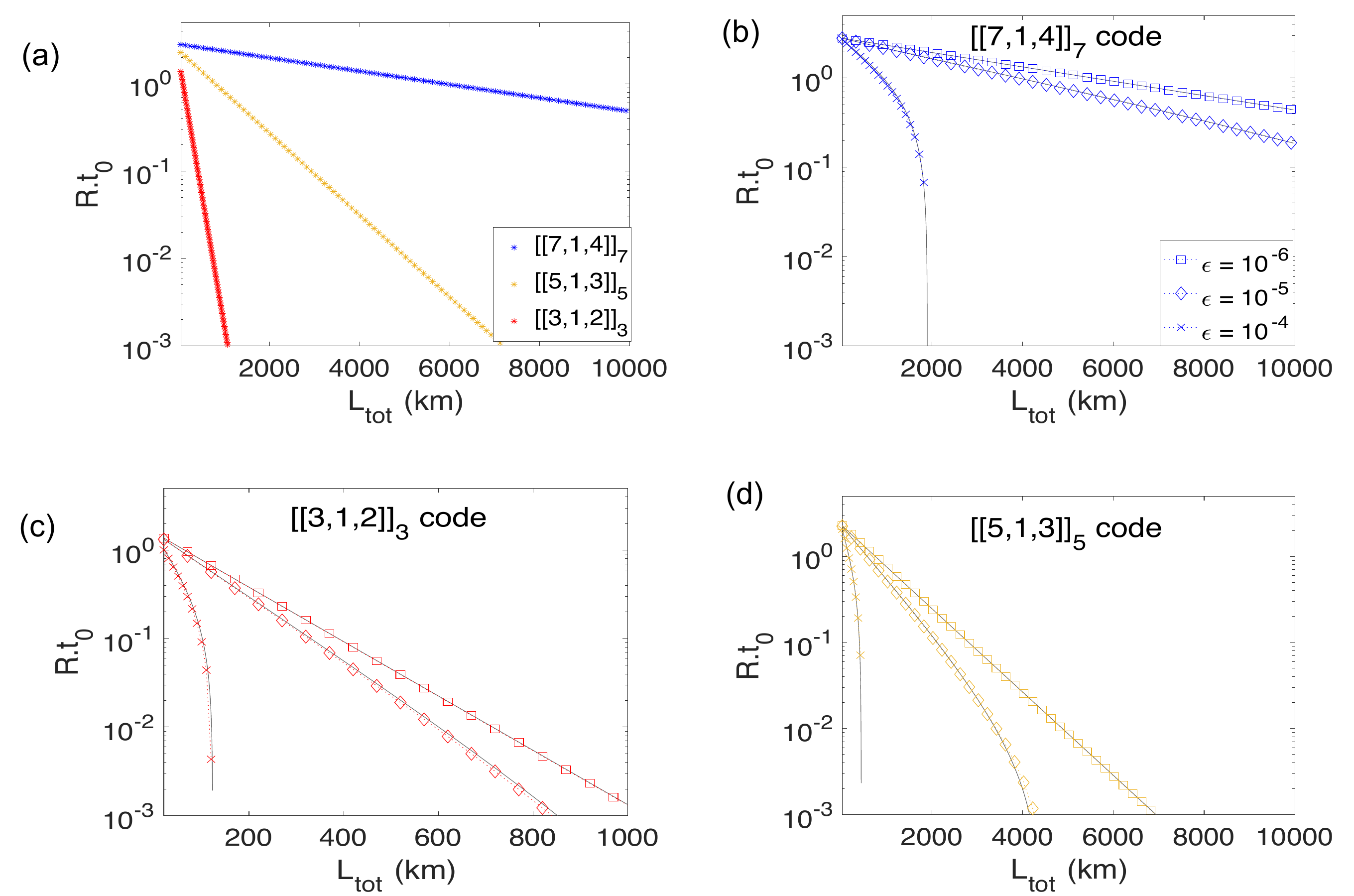}
\caption[fig:fa1]{ (a) The key generation rates $R t_0$ that can be achieved with small blocks of QPyC  in the absence of operation errors with 1km repeater spacing between the repeater stations with $[[7,1,4]]_7$ (red), $[[5,1,3]]_5$ (blue) and $[[3,1,2]]_3$ (green) codes. (b-d) $R. t_0$ in the presence of operation errors $\epsilon_g=\epsilon_d=\epsilon_p=\epsilon= 10^{-6},10^{-5},10^{-4}$ for $[[7,1,4]]_7$ (b),  $[[3,1,2]]_3$ (c) and $[[5,1,3]]_5$ (d) codes. The gray line corresponds to the  approximation taking only the leading terms into account.}
\end{center}
\label{fig:plot655}
\end{figure}

\subsection{Comparison with QPC}
We can compare the performance of different quantum codes for one-way quantum repeaters
by considering both the qubit and temporal resources consumed by the code, respectively. In order to compare QPyC codes, which use multilevel systems, with QPC of qubits, we consider 
the conversion of a $d$-level qudit into $\lceil\text{log}_2 d\rceil$ qubits and compare the performance of the codes by a cost coefficient \cite{Muralidharan14}
\begin{equation}
C'_q = \min_{k,L_{0}}\frac{2(2k+1) \lceil {\mbox{log}_2 d \rceil} } {L_0 R} \mathrm{qubits/km/sbit/s},
\end{equation}
The $C'_q$ is obtained from the product of qubit resources and temporal resources. The number of qubits used for TEC at every repeater station is given by $2(2k+1)\lceil\text{log}_2 d\rceil$ and the number of qubits for all stations is $2(2k+1) \lceil\text{log}_2 d\rceil(L_{tot}/L_{0})$ . The temporal resource used by the scheme is simply the inverse of key generation rate, i.e. $(1/R)$. Since, the product of qubit and temporal resources varies at least linearly with $L_{tot}$, we further divide the product by $L_{tot}$ to obtain the cost coefficient, which stands for the number of qubits required per km for the generation of one secure bit in one second. 

\begin{figure}[h]
\begin{center}
\includegraphics[width=5in,angle=0]{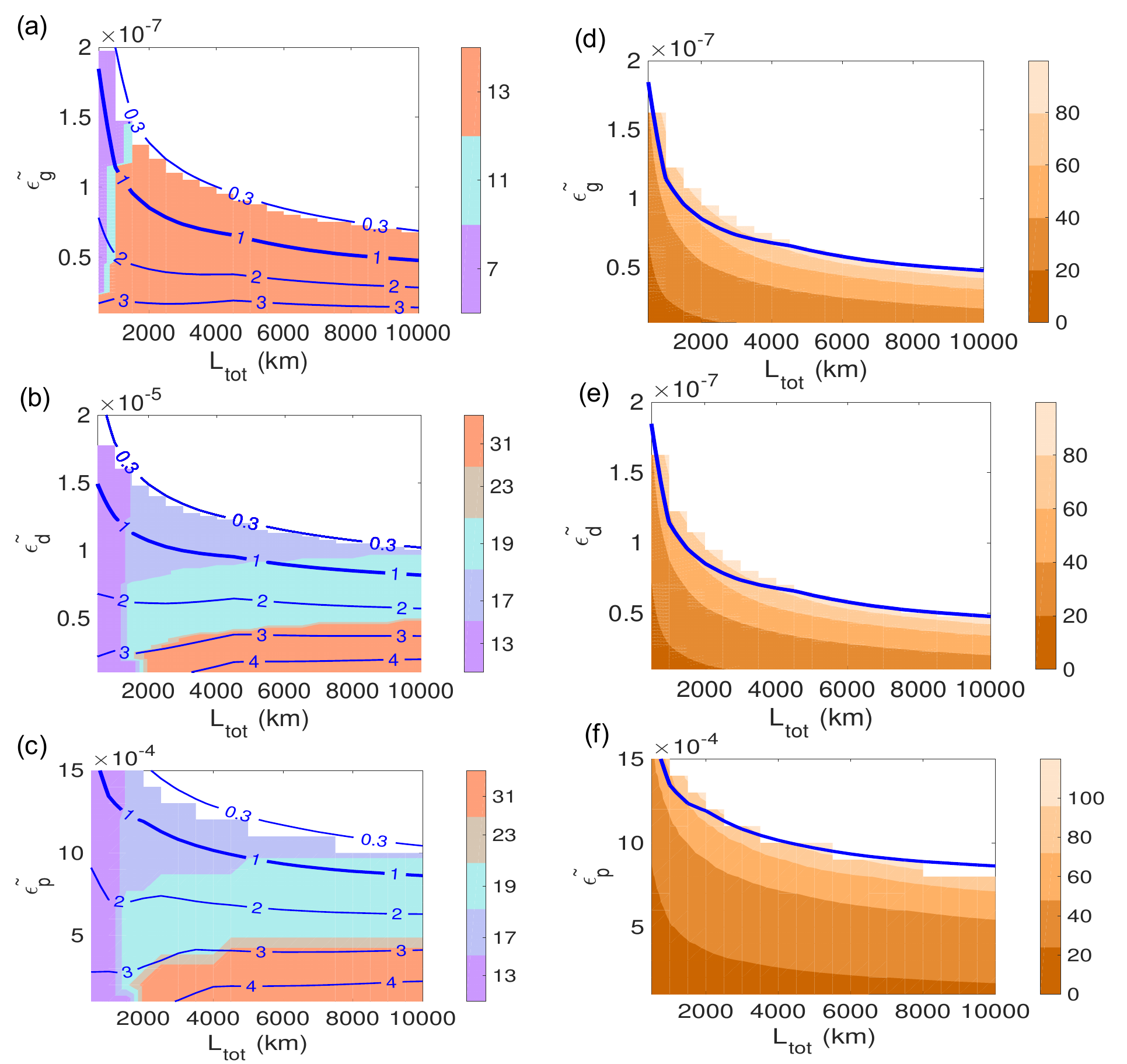}
\caption{ Contour plots (a), (b) and (c) show the improvement factor of QPyC with respect to QPC
$C'_{\mathrm{QPC}}/C'_{\mathrm{QPyC}}$ when the imperfections are dominated by 
the two-qudit gate error $\widetilde{\epsilon_g}= \epsilon_g/d^4$,  qudit depolarization error $\widetilde{\epsilon_d}= \epsilon_d/d^2$ and qudit dephasing error $\widetilde{\epsilon_p}= \epsilon_p/d$ respectively. The optimal $d$ chosen by the optimization (of $C'$) is shown with different colors in the same plots. The contour area below the break even contour line (shown with thicker line) indicates the area where QPyC performs better than QPC. Contour plots (d-f) show the variation of optimized cost coefficients $C'_q$ for different dominant errors. Here, it is assumed that it takes the same time to create small encoded block of qubits(qudits).}
\label{fig:operations}
\end{center}
\end{figure}

We find that the present scheme with QPyC can achieve a very small cost coefficient, which is about 5 times less than for QPC with TEC for $L_{\mathrm{tot}}=10,000$ km in the absence of operation errors. Note that in addition to the local resource overhead, we can also save the number of modes considerably by using QPyC instead of QPC (See Appendix A). 
In the presence of operation errors, the comparison between qubit and qudit based schemes for QRs depends largely on the error model because of the complexity involved in the implementation of multimode operations. Although it is challenging to compare the resource requirements between these codes, we have attempted to ensure a fair comparison by assuming that the operation errors increase with dimension $d$ of the qudit. We assume that
$\epsilon_g =  \widetilde {\epsilon_g}  \times d^4$, $\epsilon_d =  \widetilde {\epsilon_d}  \times d^2$, 
and $\epsilon_p =  \widetilde {\epsilon_p}  \times d$  for the comparison.  In Fig. \ref{fig:operations}, we investigate the variation of the ratio of cost coefficients $C'_{\mathrm{QPC}}/C'_{\mathrm{QPyC}}$ with respect to total distance of communication $L_{tot}$. We consider only one of the errors $\widetilde{\epsilon_g}$, $\widetilde{\epsilon_d}$ and $\widetilde{\epsilon_p}$ respectively in each one of the plots and show the variation of the ratio of the cost coefficients with respect to $L_{tot}$ and the corresponding $d$ picked by the optimization of the cost coefficient shown in the colored contour. As the  operation error increases, the ratio decreases and QPC becomes more favorable than QPyC
shown in the area above the break-even contour line (shown as thickened line) as the number of levels $d$ increases. Since $\epsilon_g$ scales as $d^4$ and  $d>2k+1$,  the optimization is forced to choose a smaller code for large total distances in  Fig. 5(a) compared to Fig. 5(b) and Fig. 5(c).
\section{Physical implementation}
For the implementation of quantum repeaters with QPyC encoding schemes,
we can consider time-bin photonic qudits. The time-bin qudit has a single
photon excitation in a superposition of $d$ time-bins, which can be efficiently generated at telecom wavelengths and coupled into an optical fiber \cite{Marcikic2004}. The TEC circuit, which is essential of our scheme, requires single qudit $X$- and $Z$-operation, two-qudit SUM gate, $X$- and $Z$-measurement of qudits. The SUM gate can be further decomposed using Controlled Z (CZ) gate and Fourier gates \cite{Gottesman1999}. 

For the time-bin qudits, the $Z$-operation and Z-measurement can be
 implemented by a selectively phase shifting or detection at different
time slots. $X$-operation and $X$-measurement can be achieved by making a strong dispersive
medium that interferes different time-bins as detailed below. For the two-qudit gate, the direct photon-photon
interaction is negligible due to weak single photon nonlinearity. Therefore, we
propose to use a nanophotonic-atom interference to store and manipulate the photonic
qudits and to mediate photon-photon qudits gate. 

The feasible experimental system adapted from Ref. \cite{Tiecke14} is illustrated in Fig. 6(a). A single atom is trapped in the
vicinity of a fiber integrated photonic crystal nanocavity, single photons
can be efficiently coupled to the nanocavity using the tapered fiber,
which strongly interacts with the atom due to the very small mode volume
of the nanocavity. 

Based on the nanophotonic atom interface, we can implement the Fourier
($F$) \cite{Gottesman1999} and $X$-gates of single time-bin qudit
by downloading the photonic time-bin qudit to the atom, and using photonic-atomic
qudit control phase (CZ) gate. Based on these elementary gates, the $X$-measurement
can be realized by the combination of $F$-gate and $Z$-measurement. 
The photon-photon qudit SUM gate can be realized by a sequence
of Fourier gate on photon qudits and two CZ gates between photon and
atom. 
In the following, we provide the details about experimental implementation
of the above element gates: (1) $F$ and $X$ gate of single qudit,
(2) the photon-atom qudit CZ gate, (3) the atom mediated photon-photon
gates between time-bin qudits. Based on these three building blocks, QPyC can be corrected
for erasure and operation errors using TEC and retransmitted to the
neighboring station. 

To illustrate the procedure, we consider the simplest case of qutrits. We choose the $D_{2}$ lines of natural $^{87}Rb$ atoms
\cite{Thompson2013}, with the transitions between hyperfine levels $\left|S_{0,1,2}\right\rangle =|5^{2}\mathrm{S}_{1/2},\ F=1,\,m_{F}=-1,0,1\rangle$
and $|5^{2}\mathrm{P}_{3/2},\ F=2,\,m_{F}=0,1,2\rangle$. In addition, we use the $\left|A\right\rangle =|5^{2}\mathrm{S}_{1/2},\ F=2,\,m_{F}=-2\rangle$
as an ancillary state for processing the photonic qutrit. 
As illustrated in Fig. 6(b), we apply a tunable external magnetic field to introduce a relative
frequency shift depending on hyperfine levels, which will be useful for the atom-photon coupling gate.

\subsection{Single qudit $F$ and $X$ gates}

Since it is difficult to make an $X$-measurement on a time-bin photonic qutrit, we propose to 
transfer the quantum state from the time-bin photonic qutrit to either a multilevel atom or multiple two-level atoms. First of all, with microwave driving of transitions between
F=1 and F=2 levels and also optical pumping, the single $^{87}Rb$ atom
can be initialized to the ancillary state. Then, for single photon
input to the cavity, it will be largely detuned from the transitions
of the atoms. By applying an external laser driving with frequency
$\omega_{R}-\omega_{s}=\Delta_R$, where $\Delta_R$ is the energy difference
between $\left|A\right\rangle $ and $\left|S_{1}\right\rangle $, through a stimulated Raman adiabatic passage \cite{Bergmann2015}, the population will
be transferred to $\left|S_{1}\right\rangle $ if there is a single
photon in the incoming time-bin, as shown in Fig. 6(b). By a sequence of microwave
pulses, the population will be mapped to $\left|S_{j}\right\rangle $
according to the order of input time-bin. This process effectively maps
the state of the time-bin qutrit to the atom's ground state levels. After that, we can carry out an $X$-measurement by performing a Fourier transformation
followed by a $Z$ measurement on the atom. Alternatively,
it is also possible to store the time-bin qutrit into multiple atoms
and control the process by synchronized pulse sequences on individual
atoms. 

This procedure can be generalized to time-bin qudits $(d>3)$ in a straightforward fashion.
\begin{figure}[h]
\begin{center}
\includegraphics[width=12cm]{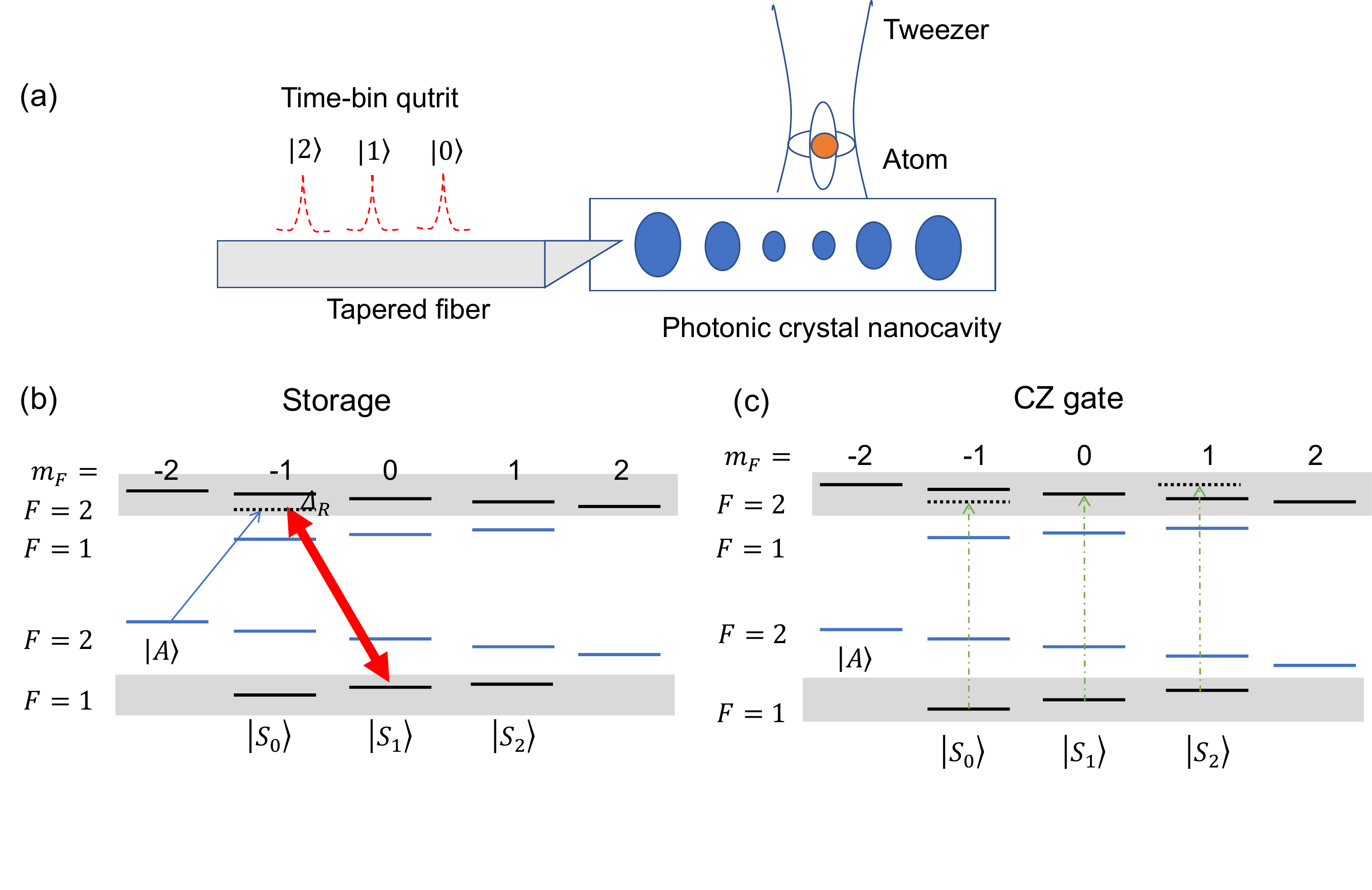}
\caption{ a) The schematic of a single atom trapped close to a photonic crystal nanocavity and the time-bin photonic qutrit input from the fiber interacting with the atom. b) The energy diagram of transfer of the quantum state of a time-bin qutrit to an atom. The atom state is initialized to the $|A\rangle$, the input photon and the strong driving (Red arrow) are detuned from the excited state by $\Delta_R$. Laser pulse followed by a joint microwave pulse is applied to map the state of the time-bin qutrit to the atom. c) The energy diagram of the CZ gate between photonic qutrit and atom.}
\label{fig:atomc}
\end{center}
\end{figure}
When the time-bin qudits are stored in atoms, arbitrary single qudit operations can be realized by applying either optical or microwave pulses. In the most general case, arbitrary unitary on single $d-$level system can be constructed by a series of $SU(2)$ gates between any two of $d$ levels (called Givens rotations) \cite{Bullock2005}. Alternatively, the arbitrary unitary can be realized with optimal controlled pulse sequences. For the qutrit case, there are more efficient schemes that only requires three steps, by either three Givens rotation or three Houlsehould's reflections \cite{Vitanov2012}.  Recently, arbitrary unitary on $d=16$ hyperfine energy levels of $^{133}Cs$ atom has been demonstrated \cite{Anderson2015}. Two-qudit gates between atoms can be mediated by cavity photons. Generalization of such techniques \cite{Vitanov2012} will enable the generation of the encoded Bell states needed for the TEC at repeater stations.

\subsection{Controlled-Z gate between photonic and atomic qudits}
To realize CZ gate between the time-bin photonic qutrit and the atom,
we prepare the atom in the superposition of $|S_{i}\rangle$. Due
to the strong cooperative interaction between the atom and nanocavity
photon, the incoming single photon pulse input to the system have
the reflectivity as \cite{Tiecke14}
\begin{equation}
r=\frac{\eta-1+2i\delta/\gamma}{\eta+1-2i\delta/\gamma},
\end{equation}
where $\eta$ is the atom-photon interaction cooperativity, $\delta$
is the frequency detuning between the pulse and atomic transition
frequency and $\gamma$ is the excited state decay rate. For a large $\eta (\approx 100)$,
$r\approx e^{i\phi}$ and $\phi\approx2\arctan\frac{2\delta}{\eta\gamma}$.
Due to the magnetic field, the photon detuning $\delta_{j}$ for different
$\left|S_{j}\right\rangle $ gradually reduces with $j$, corresponding
to a atomic state dependent phase shift as shown in Fig. 6(c). 

The CZ gate can be represented as $C=\sum_{m}|m\rangle\langle m|\otimes U_{a,m}$,
where $m=0,1,2$ denoting the states of photonic qutrits and $U_{a,m}$
is photon state dependent phase gate. The gate can be realized by
the following three steps: 
\begin{enumerate}
\item For the first incoming pulse, all energy levels are detuned from the
cavity mode by controlling external bias magnetic field such that $\delta_{0}=\delta_{1} =\delta_{2}$
then we have an operation on the atomic energy levels as $U_{a,0}=diag\{1,1,1\}.$
\item For the second pulse, magnetic field is tuned to have $\delta_{0}=\sqrt{3}\gamma\eta/2$,
$\delta_{1}=0$ and $\delta_{2}=-\sqrt{3}\gamma\eta/2$, so that the phase
shift is $2\pi/3$, $0$ and $-2\pi/3$, $U_{a,1}=diag\{e^{i\frac{2\pi}{3}},1,e^{-i\frac{2\pi}{3}}\}.$
\item For the third pulse, the magnetic field is reversed, so that the phase
shift is $-2\pi/3$, $0$ and $2\pi/3$, then $U_{a,2}=diag\{e^{-i\frac{2\pi}{3}},1,e^{i\frac{2\pi}{3}}\}.$ 
\end{enumerate}

The generalization of the CZ gate to higher dimensional qudits $(d>3)$ is less straightforward compared to time-bin storage. One possible realization
is using the cavity coupled multiple $\Lambda$-type atoms (each atom is an effective two-level system). Suppose there are $d$ atoms and each has ground
states $\left|g\right\rangle $, $\left|s\right\rangle $ and excited
state $\left|e\right\rangle $, the transition $\left|g\right\rangle \leftrightarrow\left|e\right\rangle $
is near resonance with cavity while $\left|s\right\rangle \leftrightarrow\left|e\right\rangle $
is far off-resonance. In addition, the transition frequencies of atoms
can be controlled by external electric or magnetic field individually.
First, the atoms are initialized to $\left|ss\ldots s\right\rangle $,
and transition frequencies are on-resonance with cavity. By single
photon Raman transition of the atom ensemble, the states of atoms
are prepared to the one-excitation Dicke state $\frac{1}{\sqrt{d}}(\left|gs\ldots s\right\rangle +\left|sg\ldots s\right\rangle +\ldots\left|ss\ldots g\right\rangle )$.
Then, the CZ gate of $d$-level time-bin qudit can be realized by
shifting the transition frequencies of individual atoms: for $i$-th ($i=1,\ldots,d$)
pulse input to the system, the $j$-th ($j=1,\ldots,d$) atom is
tuned to be near-resonance on the cavity so that the reflected pulse
gains a phase of $e^{-2i\frac{\pi}{d}(i-1)(j-1)}$.

\subsection{Controlled-Z gate between time-bin qudits}
A  CZ gate can be realized between two physical time-bin photonic qudits by a generalization of the Duan-Kimble scheme \cite{Duan2004}. The CZ gate between two time-bin photonic qutrits ($f$ and $s$) can be realized by preparing the atomic state in a equal superposition of all energy levels and applying three photon-atom CZ gates ($C_{f,s}=\sum_{m} |m\rangle_{f,s} \langle m| \otimes U_{a,m}$ with $m=0,1,2$ denoting the states of photonic qutrits) and two Fourier gates ($F$) on atoms as \cite{Aharonov2008}
\begin{eqnarray}
U_{CZ}= &{C_{f}}^{-1}F^{-1} {C_{s}}^{-1} F C_{f},
\label{eq:fourier}
\end{eqnarray}
where,
\begin{equation}
F=\frac{1}{\sqrt{3}}\left(\begin{array}{ccc}
1 & 1 & 1\\
1 & e^{i\frac{2\pi}{3}} & e^{i\frac{4\pi}{3}}\\
1 & e^{i\frac{4\pi}{3}} & e^{i\frac{8\pi}{3}}
\end{array}\right).
\end{equation}
The atomic state initialization and Fourier gates of atoms can be realized by microwave or optical Raman pulsed controlled transitions ($\left|S_0\right\rangle \leftrightarrow\left|S_1\right\rangle $ and $\left|S_1\right\rangle \leftrightarrow\left|S_2\right\rangle $) and phase gate $Z$.

For qudits with $d>3$, the Fourier gate of atoms can be realized by virtual cavity photon mediated atom-atom interaction, where strong pumping is applied on the $\left|s\right\rangle \leftrightarrow\left|e\right\rangle $ transition, with the detuning equals to that between cavity
photon and $\left|g\right\rangle \leftrightarrow\left|e\right\rangle $.
Based on this procedure and the CZ gate between photonic and atomic qudits $(d>3)$ described in the previous section, the CZ gate between two time-bin qudits can be realized using a similar technique as Eq. (\ref{eq:fourier}).

\section{Conclusion}
We have investigated efficient codes using multilevel systems that can correct up to $50\%$ erasure error rates, which is the bound set by the no-cloning theorem \cite{Bennett1997}. 
The success probability of quantum polynomial codes close to $50\%$ erasure rates is higher than the success probability of surface codes of all code sizes. We employed teleportation based error correction that can correct erasure errors up to the threshold efficiently and discussed its application for the construction of highly efficient one-way quantum repeater networks. In comparison with quantum parity codes, we have obtained an improvement in resources by about 5 times for communication across 10,000km in the absence of operation errors and numerically identified the parameter regime, where the quantum polynomial codes perform better than quantum parity codes in the presence of operation errors. We have discussed the physical implementation of quantum polynomial codes and identified the key technological requirements.
It will also be interesting to consider the experimental implementation of multilevel systems using oscillators \cite{Albert2015}, Rydberg atoms \cite{Haroche2000} and ensembles of multilevel systems \cite{Brion2008}. 
Moreover, we can extend the coding schemes to continuous variables for the correction of erasure errors \cite{Li2016, Bergmann2016}. Besides quantum communication, quantum error correcting codes for multilevel systems that can correct a large fraction of erasure errors might be useful for improving
precision metrology \cite{sense1, sense2, sense3, preskillclock, pleniobook}. 

\section*{Acknowledgements}
This work was supported by the DARPA (Quiness program), ARL CDQI, NBRPC (973 program), Packard Foundation, Alfred P. Sloan Foundation, ARO (W911NF-14-1-0011, W911NF-14-1-0563), ARO MURI (W911NF-16-1-0349), NSF (EFMA-1640959), AFOSR MURI (FA9550-14-1-0052, FA9550-15-1-0015). We thank Siddharth Prabhu, Anup Rao, Victor Albert, Jungsang Kim, Norbert L\"utkenhaus, Mikhail Lukin, Hong Tang and Steven Girvin for discussions.

\section*{APPENDIX A}
\begin{figure}[h]
\begin{center}
\includegraphics[width=8cm]{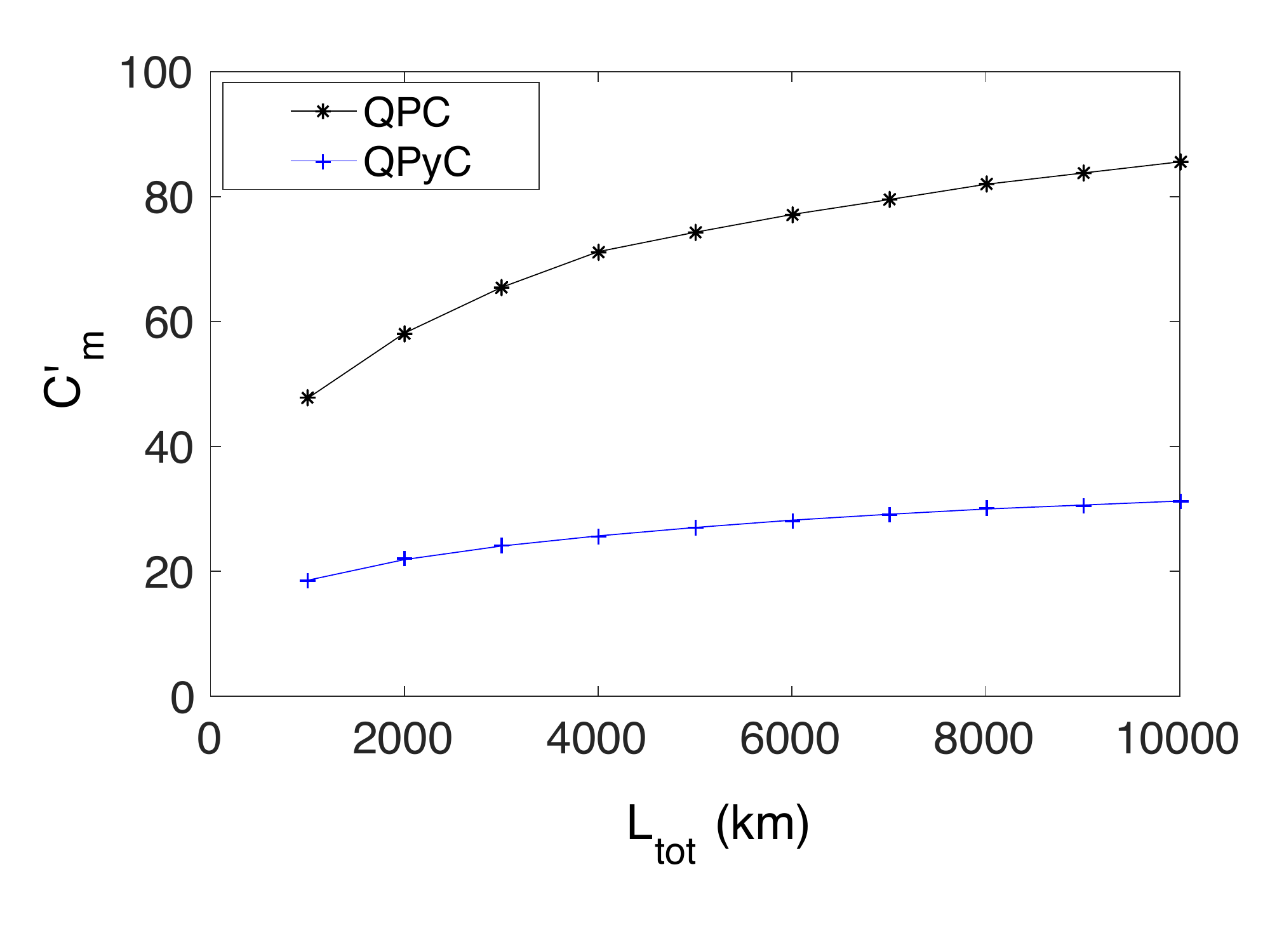}
\caption[fig:fa1]{The cost coefficient $C'_m$ based on the number of modes for QPC and QPyC.}
\label{fig:qpccostmode2}
\end{center}
\end{figure}

In the manuscript, we have assumed that we are only limited by the qubit resources rather than the
optical modes of the fiber channel. If qubit resource is no longer the limiting factor for one-way quantum repeaters, we should then compare the number of modes needed for different repeater schemes. 
For quantum state transfer using time-bin qudits, each qudit needs $d$ temporal modes for transmission. Therefore, the cost coefficient based on the number of modes for the transmission of photons is given by,
\begin{equation}
C'_m(L_{tot}) = \min_{k,L_{0}}\frac{(2k+1)d } {L_0 R} \mathrm{modes/km/sbit/s}.
\end{equation}
In Fig. \ref{fig:qpccostmode2}, we compare the cost coefficients $C'_m$ of QPyC and QPC. For QPC, since each qubit needs two temporal modes for transmission the numerical values of $C'_q$ and $C'_m$ are the same. For QPyC, the numerical value of $C'_m$ is increased by a small factor compared to $C'_q$. We find that for $L_{\mathrm{tot}}=10,000$ km, QPyC achieves a cost that is about 3 times less than QPC. 
\bibliographystyle{iopart-num}
\def\urlprefix{}
\def\url#1{}
\bibliography{quditcitations.bib}

\providecommand{\newblock}{}
\begin{thebibliography}{10}
\expandafter\ifx\csname url\endcsname\relax
  \def\url#1{{\tt #1}}\fi
\expandafter\ifx\csname urlprefix\endcsname\relax\def\urlprefix{URL }\fi
\providecommand{\eprint}[2][]{\url{#2}}

\bibitem{Bennett1997}
Bennett C~H, DiVincenzo D~P and Smolin J~A 1997 {\em Phys. Rev. Lett.\/} {\bf
  78} 3217

\bibitem{Preskill1998}
Preskill J 1998 {\em Proc. Roy. Soc. London\/} {\bf 454} 385
  \urlprefix\url{http://rspa.royalsocietypublishing.org/cgi/doi/10.1098/rspa.1998.0167}

\bibitem{Wu2002}
Wu L~A, Byrd M~S and Lidar D~A 2002 {\em Phys. Rev. Lett.\/} {\bf 89} 127901
  \urlprefix\url{http://link.aps.org/doi/10.1103/PhysRevLett.89.127901}

\bibitem{Fazio1999}
Fazio R, Palma G~M and Siewert J 1999 {\em Phys. Rev. Lett.\/} {\bf 83} 5385
  \urlprefix\url{http://link.aps.org/doi/10.1103/PhysRevLett.83.5385}

\bibitem{Vala2005}
Vala J, Whaley K~B and Weiss D~S 2005 {\em Phys. Rev. A\/} {\bf 72} 052318
  \urlprefix\url{http://link.aps.org/doi/10.1103/PhysRevA.72.052318}

\bibitem{Knill2001}
Knill E, Laflamme R and Milburn G~J 2001 {\em Nature (London)\/} {\bf 409} 46
  \urlprefix\url{http://www.ncbi.nlm.nih.gov/pubmed/11343107}

\bibitem{Duan2004}
Duan L~M and Kimble H~J 2004 {\em Phys. Rev. Lett.\/} {\bf 92} 127902
  \urlprefix\url{http://link.aps.org/doi/10.1103/PhysRevLett.92.127902}

\bibitem{Kok2007}
Kok P, Munro W~J, Nemoto K, Ralph T~C, Dowling J~P and Milburn G~J 2007 {\em
  Rev. Mod. Phys.\/} {\bf 79} 135
  \urlprefix\url{http://link.aps.org/doi/10.1103/RevModPhys.79.135}

\bibitem{Bell2014}
Bell B~A, Herrera-Mart{\'{i}} D~A, Tame M~S, Markham D, Wadsworth W~J and
  Rarity J~G 2014 {\em Nature Comm.\/} {\bf 5} 3658
  \urlprefix\url{http://www.ncbi.nlm.nih.gov/pubmed/24752224
  http://www.nature.com/doifinder/10.1038/ncomms4658}

\bibitem{Muralidharan2015a}
Muralidharan S, Li L, Kim J, L{\"{u}}tkenhaus N, Lukin M~D and Jiang L 2016
  {\em Scientific Reports\/} {\bf 6} 20463

\bibitem{Briegel98}
Briegel H~J, Dur W, Cirac J~I and Zoller P 1998 {\em Phys. Rev. Lett.\/} {\bf
  81} 5932

\bibitem{Jiang2008}
Jiang L, Taylor J~M, Nemoto K, Munro W~J, {Van Meter} R and Lukin M~D 2009 {\em
  Phys. Rev. A\/} {\bf 79} 32325

\bibitem{Bratzik14}
Bratzik S, Kampermann H and Bruss D 2014 {\em Phys. Rev. A\/} {\bf 89} 32335

\bibitem{Epping2016}
Epping M, Kampermann H and Bru{\ss} D 2016 {\em Applied Physics B: Lasers and
  Optics\/} {\bf 122} 54

\bibitem{Pant2016}
Pant M, Krovi H, Englund D and Guha S 2016 {\em arxiv: quantph:/1603.01353\/}

\bibitem{Fowler2010}
Fowler A~G, Wang D~S, Hill C~D, Ladd T~D, {Van Meter} R and Hollenberg L~C~L
  2010 {\em Phys. Rev. Lett.\/} {\bf 104} 180503

\bibitem{Munro2012a}
Munro W~J, Stephens A~M, Devitt S~J, Harrison K~A and Nemoto K 2012 {\em Nature
  Phot.\/} {\bf 6} 777

\bibitem{Muralidharan2014}
Muralidharan S, Kim J, L{\"{u}}tkenhaus N, Lukin M~D and Jiang L 2014 {\em
  Phys. Rev. Lett.\/} {\bf 112} 250501

\bibitem{Ewert2015}
Ewert F, Bergmann M and van Loock P 2016 {\em Phys. Rev. Lett.\/} {\bf 117}
  210501

\bibitem{Namiki2016}
Namiki R, Jiang L, Kim J and L{\"{u}}tkenhaus N 2016 {\em Phys. Rev. A\/} {\bf
  94} 052304

\bibitem{glaudell2016}
Glaudell A~N, Waks E and Taylor J~M 2016 {\em New Jour. of Phys.\/} {\bf 18}
  093008
  \urlprefix\url{http://stacks.iop.org/1367-2630/18/i=9/a=093008?key=crossref.532d612b4f9628dc8b5ec1c563c8ed3c}

\bibitem{Varnava2006}
Varnava M, Browne D~E and Rudolph T 2006 {\em Phys. Rev. Lett.\/} {\bf 97}
  120501 \urlprefix\url{http://link.aps.org/doi/10.1103/PhysRevLett.97.120501}

\bibitem{Barrett2010}
Barrett S~D and Stace T~M 2010 {\em Phys. Rev. Lett.\/} {\bf 105} 200502
  \urlprefix\url{http://link.aps.org/doi/10.1103/PhysRevLett.105.200502}

\bibitem{Stace2010}
Stace T~M and Barrett S~D 2010 {\em Phys. Rev. A\/} {\bf 81} 022317
  \urlprefix\url{http://link.aps.org/doi/10.1103/PhysRevA.81.022317}

\bibitem{Gingrich2003}
Gingrich R~M, Kok P, Lee H, Vatan F and Dowling J~P 2003 {\em Phys. Rev.
  Lett.\/} {\bf 91} 217901
  \urlprefix\url{http://link.aps.org/doi/10.1103/PhysRevLett.91.217901}

\bibitem{Muralidharan14}
Muralidharan S, Kim J, L{\"{u}}tkenhaus N, Lukin M~D and Jiang L 2014 {\em
  Phys. Rev. Lett.\/} {\bf 112} 250501

\bibitem{Aharonov2008}
Aharonov D and Ben-Or M 2008 {\em SIAM J. Comput.\/} {\bf 38} 1207

\bibitem{Cleve1999}
Cleve R, Gottesman D and Lo H~K 1999 {\em Phys. Rev. Lett.\/} {\bf 83} 648
  \urlprefix\url{http://link.aps.org/doi/10.1103/PhysRevLett.83.648}

\bibitem{Lu2008}
Lu C~Y, Gao W~B, Zhang J, Zhou X~Q, Yang T and Pan J~W 2008 {\em Proc. Natl.
  Acad. Sci. USA\/} {\bf 105} 11050
  \urlprefix\url{http://www.pubmedcentral.nih.gov/articlerender.fcgi?artid=2516277{\&}tool=pmcentrez{\&}rendertype=abstract}

\bibitem{Grassl1997}
Grassl M, Beth T and Pellizzari T 1997 {\em Phys. Rev. A\/} {\bf 56} 33
  \urlprefix\url{http://link.aps.org/doi/10.1103/PhysRevA.56.33}

\bibitem{Brinew2}
Zwerger M, Briegel H~J and D{\"{u}}r W 2014 {\em Sci. Rep.\/} {\bf 4} 5364

\bibitem{Ralph2005}
Ralph T~C, Hayes A~J~F and Gilchrist A 2005 {\em Phys. Rev. Lett.\/} {\bf 95}
  100501 \urlprefix\url{http://link.aps.org/doi/10.1103/PhysRevLett.95.100501}

\bibitem{Knill2005a}
Knill E 2005 {\em Phys. Rev. A\/} {\bf 71} 042322
  \urlprefix\url{http://link.aps.org/doi/10.1103/PhysRevA.71.042322}

\bibitem{Knill2005}
Knill E 2005 {\em Nature (London)\/} {\bf 434} 39
  \urlprefix\url{http://www.ncbi.nlm.nih.gov/pubmed/15744292}

\bibitem{Gottesman1999}
Gottesman D 1999 {\em Chaos, Solitons {\&} Fract.\/} {\bf 10} 1749
  \urlprefix\url{http://linkinghub.elsevier.com/retrieve/pii/S0960077998002185}

\bibitem{Sheridan2010}
Sheridan L and Scarani V 2010 {\em Phys. Rev. A\/} {\bf 82} 030301
  \urlprefix\url{http://link.aps.org/doi/10.1103/PhysRevA.82.030301}

\bibitem{Marcikic2004}
Marcikic I, {De Riedmatten} H, Tittel W, Zbinden H, Legr{\'{e}} M and Gisin N
  2004 {\em Phys. Rev. Lett.\/} {\bf 93} 180502

\bibitem{Tiecke14}
Tiecke T~G, Thompson J~D, de~Leon N~P, Liu L~R, Vuletic V and Lukin M~D 2014
  {\em Nature(London)\/} {\bf 508} 241

\bibitem{Thompson2013}
Thompson J~D, Tiecke T~G, de~Leon N~P, Feist J, Akimov A~V, Gullans M, Zibrov
  A~S, Vuleti{\'{c}} V and Lukin M~D 2013 {\em Science\/} {\bf 340} 1202
  \urlprefix\url{http://www.ncbi.nlm.nih.gov/pubmed/23618764}

\bibitem{Bergmann2015}
Bergmann K, Vitanov N~V and Shore B~W 2015 {\em Jour. of Chem. Phys.\/} {\bf
  142} 170901 \urlprefix\url{http://dx.doi.org/10.1063/1.4916903}

\bibitem{Bullock2005}
Bullock S~S, O'Leary D~P and Brennen G~K 2005 {\em Phys. Rev. Lett.\/} {\bf 94}
  230502

\bibitem{Vitanov2012}
Vitanov N~V 2012 {\em Phys. Rev. A\/} {\bf 85} 032331

\bibitem{Anderson2015}
Anderson B~E, Sosa-Martinez H, Riofr{\'{i}}o C~A, Deutsch I~H and Jessen P~S
  2015 {\em Phys. Rev. Lett.\/} {\bf 114} 240401
  \urlprefix\url{http://link.aps.org/doi/10.1103/PhysRevLett.114.240401}

\bibitem{Albert2015}
Albert V~V, Shu C, Krastanov S, Shen C, Liu R~B, Yang Z~B, Schoelkopf R~J,
  Mirrahimi M, Devoret M~H and Jiang L 2016 {\em Phys. Rev. Lett.\/} {\bf 116}
  140502

\bibitem{Haroche2000}
Rauschenbeutel A, Nogues G and Osnaghi S 2000 {\em Science\/} {\bf 288} 2024
  \urlprefix\url{http://www.sciencemag.org/content/288/5473/2024.short}

\bibitem{Brion2008}
Brion E, Pedersen L~H, Saffman M and M{\o}lmer K 2008 {\em Phys. Rev. Lett.\/}
  {\bf 100} 110506
  \urlprefix\url{http://link.aps.org/doi/10.1103/PhysRevLett.100.110506}

\bibitem{Li2016}
Li L, Zou C~L, Albert V~V, Muralidharan S, Girvin S~M and Jiang L 2016 {\em
  arXiv:1609.06386v1\/} \urlprefix\url{http://arxiv.org/abs/1609.06386}

\bibitem{Bergmann2016}
Bergmann M and {Van Loock} P 2016 {\em Phys. Rev. A\/} {\bf 94} 012311

\bibitem{sense1}
Kessler E~M, Lovchinsky I, Sushkov A~O and Lukin M~D 2014 {\em Phys. Rev.
  Lett.\/} {\bf 112} 150802
  \urlprefix\url{http://link.aps.org/doi/10.1103/PhysRevLett.112.150802}

\bibitem{sense2}
Arrad G, Vinkler Y, Aharonov D and Retzker A 2014 {\em Phys. Rev. Lett.\/} {\bf
  112} 150801
  \urlprefix\url{http://link.aps.org/doi/10.1103/PhysRevLett.112.150801}

\bibitem{sense3}
D{\"{u}}r W, Skotiniotis M, Fr{\"{o}}wis F and Kraus B 2014 {\em Phys. Rev.
  Lett.\/} {\bf 112} 080801
  \urlprefix\url{http://link.aps.org/doi/10.1103/PhysRevLett.112.080801}

\bibitem{preskillclock}
Preskill J 2000 {\em arXiv:quant-ph/0010098\/}
  \urlprefix\url{http://arxiv.org/abs/quant-ph/0010098}

\bibitem{pleniobook}
Macchiavello C, Huelga S~F, Cirac J~I, Ekert A~K and Plenio M~B 1998 {\em
  Quantum Comm. Comput., and Measurement\/}  337
  \urlprefix\url{http://link.springer.com/chapter/10.1007/0-306-47097-7{\_}45}

\end{thebibliography}
\end{document}